\documentclass[superscriptaddress,nofootinbib,amsmath,floatfix,amssymb,prd,showpacs,twocolumn]{revtex4-2}
\usepackage{siunitx}
\usepackage{nicefrac}
\usepackage{graphicx}
\usepackage{float,slashed}
\usepackage[caption=false]{subfig}
\usepackage{xcolor}

\usepackage[colorlinks]{hyperref}
\hypersetup{linkcolor=red,citecolor=blue,urlcolor=blue}
\begin{document}
\captionsetup{justification=justified, singlelinecheck=false}

\title{Neutron stars with crossover to color superconducting quark matter
}
\author{D.~Blaschke}
\email{david.blaschke@uwr.edu.pl}
\affiliation{Institute of Theoretical Physics,
    University of Wroclaw,
    50-204 Wroclaw, Poland}
\affiliation{Bogoliubov Laboratory of Theoretical Physics,
    Joint Institute for Nuclear Research,
    141980 Dubna, Russia}
\affiliation{National Research Nuclear University (MEPhI),
    115409 Moscow, Russia}
\author{E.-O. Hanu}
\email{hanu@theor.jinr.ru}
\affiliation{Faculty of Physics,
    University of Bucharest,
    Bucharest, Romania}
\affiliation{Bogoliubov Laboratory of Theoretical Physics,
    Joint Institute for Nuclear Research,
    141980 Dubna, Russia}
\author{S.~Liebing}
\email{science@liebing.cc}
\affiliation{Bogoliubov Laboratory of Theoretical Physics,
    Joint Institute for Nuclear Research,
    141980 Dubna, Russia}

\date{\today}
\begin{abstract}
We follow the idea that the QCD phase diagram may be described by a crossover from a hadron resonance gas to perturbative QCD using the switch function ansatz of Albright, Kapusta and Young \cite{Albright:2014gva}.
While the switch function could be calibrated at vanishing baryon chemical potential with data from lattice QCD simulations, it has been suggested recently by Kapusta and Welle \cite{Kapusta:2021ney} that in the zero temperature limit, the switch function parameter $\mu_0$ could be constrained by neutron star phenomenology, in particular by massive pulsars like PSR J0740+6620 with a mass exceeding $2~M_\odot$.
In this work we demonstrate that this procedure to constrain the QCD phase diagram does crucially depend on the fact that cold dense quark matter is very likely in a color superconducting state.
\end{abstract}

     
\maketitle

\section{Introduction}

The quest for the structure of the phase diagram of quantum chromodynamics (QCD) in the plane of temperature and baryon density is one of the great challenges in experimental and theoretical particle and nuclear physics.
Lattice gauge theory simulations of QCD at vanishing and small net baryon densites have revealed that the transition from hadronic matter to the deconfined quark-gluon plasma (QGP) is a crossover \cite{HotQCD:2018pds}. 
Searches for the critical endpoint (CEP) of a possible first order phase transition at higher baryon densities and lower temperatures, where lattice QCD could not reach at present due to the sign problem, have been performed experimentally with the beam energy scan (BES) programs of ultrarelativistic heavy-ion collisions.
In particular, the STAR experiment at RHIC Brookhaven could not provide conclusive evidence for a CEP from their phase one of the BES program. 
It has been pointed out recently (see \cite{Senger:2021dot} and references therein) that in order to reach the CEP  in a heavy-ion collision, the center of mass energies must be below $\sqrt{s_{NN}}\approx$\,\SI{6}{GeV}, because the temperature of the CEP should not exceed that of the chiral transition in the chiral limit which has been determined by lattice QCD to be $T_c^0=132^{+3}_{-6}$ MeV \cite{HotQCD:2019xnw}.
On the other hand, at low temperatures the transition is also likely to be a crossover, due to the possible coexistence of chiral symmetry breaking and color superconductivity which is induced by the mixing of the corresponding order parameters by the Fierz-transformed Kobayashi-Maskawa-'t Hooft determinant interaction resulting in a second CEP \cite{Hatsuda:2006ps} or even the absence of a CEP at all. 
Such a situation would be in accordance with the concept of hadron-quark continuity \cite{Schafer:1998ef,Schafer:1999pb,Yamamoto:2007jn}.

In this situation that the structure of the QCD phase diagram is likely to be a "crossover all over", an approach to a unified description of the equation of state (EoS) of hadronic and quark matter phases has been suggested \cite{Albright:2014gva} which is based on an interpolation between a hadron resonance gas EoS and a perturbative QCD approach to the QGP using a switching function
\begin{equation}
\label{eq:switch}
    S(T,\mu)=\exp\left[-(T_0/T)^q-(\mu_0/\mu)^r\right].
\end{equation}
Here $T$ is the temperature and $\mu$ the baryochemical potential.
The exponents $q$ and $r$ are free parameters for which in \cite{Albright:2014gva} it has been assumed that $r=q=4, 5$.
The temperature $T_0$ sets the scale for the transition on the temperature axis and has been calibrated by comparison with the EoS of lattice QCD thermodynamics.
In order to fix the scale $\mu_0$ for the transition in the direction of the baryochemical potential it has been suggested recently by Kapusta and Welle\,\cite{Kapusta:2021ney} to employ a comparison with observational data for the maximum mass of neutron stars that is uniquely determined by the EoS of nuclear matter in $\beta$-equilibrium at $T=0$.

However, in their study \cite{Kapusta:2021ney} Kapusta and Welle employed a rather rudimentary EoS for $T=0$ quark matter which in particular neglected the effects of color superconductivity that were considered essential for the emergence of a crossover transition at low temperatures. 

In the present work, we will follow the idea of Ref.~\cite{Kapusta:2021ney} to fix the switch function in the $T=0$ limit by a comparison with neutron star phenomenology.
Going beyond their setup, we will employ a color superconducting quark matter EoS with a diquark pairing gap $\Delta$, allowing also for an effective bag pressure $B_{\rm eff}$ that mimics confining effects.
As the width of the crossover transition is expected to be temperature dependent and become narrower for lower temperatures, even in the crossover-all-over scenario, we will investigate the influence of narrowing the switching function by increasing the r parameter.

We will use modern multi-messenger data from the NICER experiment \cite{Miller:2021qha} for constraining the mass-radius data of the massive pulsar PSR J0740+6620 \cite{Fonseca:2021wxt} and from the LIGO-Virgo Collaboration for the tidal deformability of a neutron star at $1.4~M_\odot$, extracted from the gravitational wave signal of the binary neutron star merger GW170817\,\cite{LIGOScientific:2018cki}.

\section{Equation of state}
\subsection{Quark matter}
We will use the form of EoS for color superconducting quark matter phases that was suggested in Alford\,\emph{et al.}\,\cite{Alford:2004pf} and recently used again in Ref.\,\cite{Zhang:2020jmb}
\begin{equation}
\label{eq:Pq1}
    P_q(\mu)=\frac{\xi_4 a_4}{4\pi^2} \left(\frac{\mu}{3}\right)^4 + \frac{\xi_{2a}\Delta^2 -\xi_{2b} m_s}{\pi^2} \left(\frac{\mu}{3}\right)^2  + \mu_e^4 - B_{\rm eff}~,
\end{equation}
where for the color-flavor-locking (CFL) phase holds that $\xi_4=$\,\num{3}, $\xi_{2a}=$\,\num{3} and $\xi_{2b}=$\,\nicefrac{3}{4}.
As suggested in \cite{Kapusta:2021ney}, we will consider massless quarks so that $m_u=m_d=m_s=$\,\num{0} and electric neutrality is manifest even without leptons, i.e. $\mu_e=$\num{0}.
For the coefficient $a_4=1-2\alpha_s/\pi$ we will use here the constant values $a_4=$\,\num{0.7} or $a_4=$\,\num{0.3} that could be attained in the nonperturbative domain relevant for hybrid neutron stars. 
The resulting three-flavor, color superconducting quark matter EoS reads
\begin{equation}
\label{eq:Pq2}
P_q(\mu)=\frac{3 }{4\pi^2} a_4 \left(\frac{\mu}{3}\right)^4 + \frac{3}{\pi^2}\Delta^2 \left(\frac{\mu}{3}\right)^2 - B_{\rm eff}~.
\end{equation}
For the quark number density follows
\begin{equation}
    n_q(\mu) = 3 \frac{\partial P_q(\mu)}{\partial \mu} 
    = \frac{3 }{\pi^2} a_4 \left(\frac{\mu}{3}\right)^3 + \frac{6}{\pi^2}\Delta^2 \left(\frac{\mu}{3}\right)~,
\end{equation}
and the energy density is thus
\begin{eqnarray}
    \varepsilon_q(\mu) &=& - P_q(\mu) + \left(\frac{\mu}{3}\right) n_q(\mu)
    \nonumber\\
       &=& \frac{9}{4 \pi^2} a_4 \left(\frac{\mu}{3}\right)^4 + \frac{3}{\pi^2}\Delta^2 \left(\frac{\mu}{3}\right)^2+ B_{\rm eff}~.
\end{eqnarray}
An interesting quantity is the squared sound speed which serves as a measure for the stiffness of the EoS.
It is obtained as
\begin{eqnarray}
\label{eq:cs2}
    c_s^2(\mu) &=& \frac{d P_q(\mu)}{d \varepsilon(\mu)}
    = \frac{n_q(\mu)}{\mu d n_q(\mu)/d \mu}
    =\frac{1+\zeta(\mu)}{3+\zeta(\mu)}~,
\end{eqnarray}
where $\zeta(\mu)=2(3\Delta)^2/(a_4 \mu^2)$. We like to discuss two limits. 

\begin{figure*}[!htb]
\centering
 \subfloat{\hspace{-2em}
\includegraphics[width=0.33\textwidth]{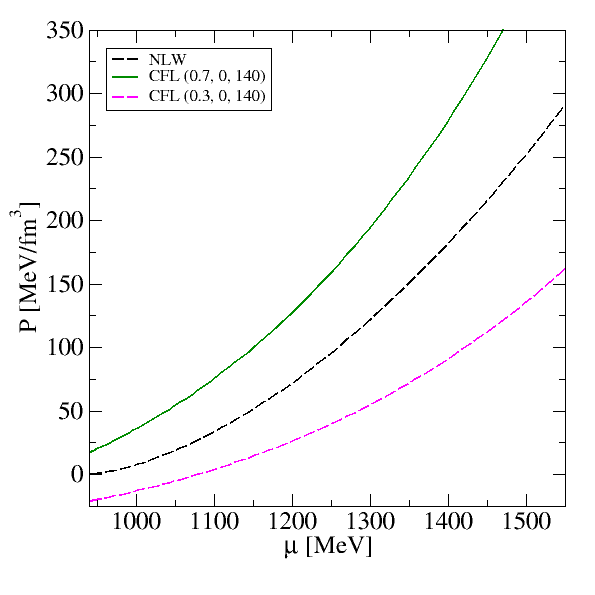}\hspace{-0.8em}
\includegraphics[width=0.33\textwidth]{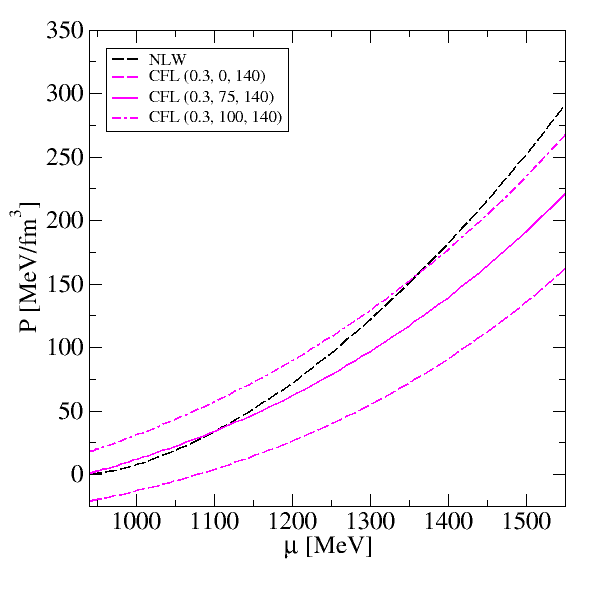}\hspace{-0.8em}
\includegraphics[width=0.33\textwidth]{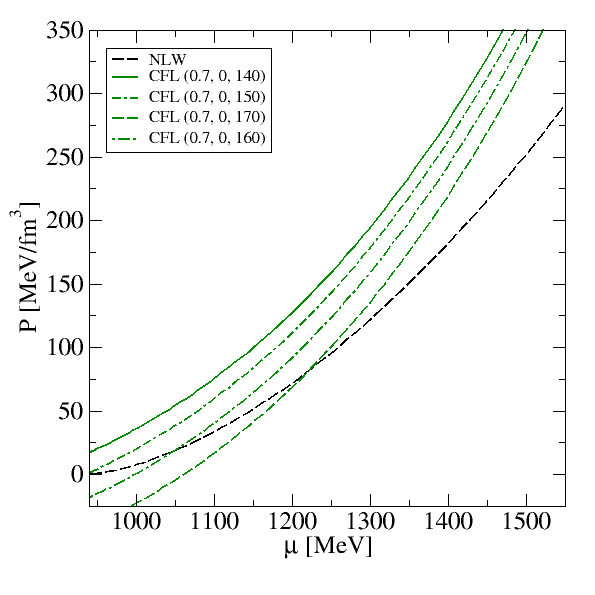}}
\vspace{-5mm}
\caption{The influence of the parameters in the EoS (\ref{eq:Pq2}) for the quark pressure as a function of the baryochemical potential: the $a_4$ parameter (left panel), the diquark pairing gap (middle panel) and the bag constant (right panel). The quark matter models are defined by the three parameters in round brackets ($a_4$, $\Delta$[MeV], $B_{\rm eff}^{1/4}$[MeV]). For comparison, the pressure of pure neutron matter in the nonlinear Walecka (NLW) model is shown by the black dashed line.}
\label{fig:P_mu_plots}
\end{figure*}

For normal quark matter, when $\Delta=$\,\num{0}, the squared sound speed obeys the "conformal limit" case $c_s^2=$\,\nicefrac{1}{3}. 
Immediately after the deconfinement transition, when $\mu\approx \mu_c \approx$\,\SI{1150}{MeV} and for large diquark pairing gap, $\Delta\approx$\,\SI{150}{MeV}, the parameter $\zeta(\mu_c) \approx$\,\num{1} may be attained for $a_4=$\,\num{0.3} so that  $c_s^2(\mu_c) =$\,\nicefrac{1}{2}. 
This value has been obtained as a typical result for several parametrizations of a 
instantaneous nonlocal chiral quark model\,\cite{Antic:2021zbn}.
Within an instantaneous separable parametrization of the nonlocal chiral quark model that fits the three-momentum dependence of the quark mass function obtained in Coulomb gauge lattice QCD it has recently been demonstrated \cite{Contrera:2022tqh} that the speed of sound for zero temperature quark matter in $\beta$-equilibrium  attains almost constant values as a function of the energy density, in the vicinity of 
$c_s^2=$\,\num[separate-uncertainty=true]{0.5\pm0.1} depending on the values of vector meson and diquark coupling. 
It is the main aim of this work to investigate the modifications relative to the hadron-to-quark matter crossover scenario discussed in Ref.\,\cite{Kapusta:2021ney} that result from the introduction of the color superconductivity term proportional to $\Delta^2 \mu^2$ in the equation of state.

For modeling the hadronic matter phase we use the description of pure neutron matter as given in the nonlinear Walecka model, as in Ref.~\cite{Kapusta:2021ney}. 

\subsection{Neutron matter}

The pressure and energy density of pure neutron matter in the nonlinear Walecka (NLW) model are given by\,\cite{Haensel:2007yy,Glendenning:1997wn} 
\begin{eqnarray}
\label{eq:Ph1}
P_h(n) &=&P_{\rm FG} + \frac{1}{2} \left[
\left(\frac{g_\omega}{m_\omega} \right)^2 
+ \frac{1}{4} \left(\frac{g_\rho}{m_\rho} \right)^2
\right] n^2
\nonumber\\
&-& \frac{1}{2} \left(\frac{m_\sigma}{g_\sigma} \right)^2 (g_\sigma \sigma)^2 
- \frac{1}{3} b m (g_\sigma \sigma)^3 - \frac{1}{4} c (g_\sigma \sigma)^4~, 
\nonumber\\
\\
\varepsilon_h(n) &=& \varepsilon_{\rm FG} + \frac{1}{2} \left[
\left(\frac{g_\omega}{m_\omega} \right)^2 
+ \frac{1}{4} \left(\frac{g_\rho}{m_\rho} \right)^2
\right] n^2
\nonumber\\
&+& \frac{1}{2} \left(\frac{m_\sigma}{g_\sigma} \right)^2 (g_\sigma \sigma)^2 
+ \frac{1}{3} b m (g_\sigma \sigma)^3 + \frac{1}{4} c (g_\sigma \sigma)^4~, 
\nonumber\\
\label{eq:eps}
\end{eqnarray}
where the Fermi gas pressure and energy density are analytically given as \cite{Kapusta:1989tk}
\begin{eqnarray}
\label{eq:PFG}
P_{\rm FG}&=&\frac{1}{3\pi^2}\int_0^{k_F} dk \frac{k^4}{E^*(k)}
\nonumber\\
&=&\frac{1}{8\pi^2}
\left[\frac{2}{3}E_F^*k_F^3 - {m^*}^2 E_F^*k_F 
+ {m^*}^4 \ln\left(\frac{E_F^* + k_F}{m^*} \right) \right], 
\nonumber\\
\\
\varepsilon_{\rm FG}&=&\frac{1}{\pi^2}\int_0^{k_F} dk{k^2}{E^*(k)}
\nonumber\\
&=&\frac{1}{8\pi^2}
\left[{2}{E_F^*}^3 k_F - {m^*}^2 E_F^*k_F 
- {m^*}^4 \ln\left(\frac{E_F^* + k_F}{m^*} \right) \right]. 
\nonumber\\
\label{eq:eFG}
\end{eqnarray}
The neutron Fermi momentum $k_F$ is given by the baryon density $n=k_F^3/(3\pi^2)$ in pure neutron matter, while the Fermi energy is
$E_F^*=\sqrt{k_F^2+{m^*}^2}$ with the neutron effective mass $m^*=m-g_\sigma \sigma$.
The scalar mean field is given by 
\begin{equation}
\label{eq:sigma}
    g_\sigma \sigma=({g_\sigma}/{m_\sigma})^2 
\left[n_s - b m (g_\sigma \sigma)^2 - c (g_\sigma \sigma)^3\right]~, 
\end{equation}
where the scalar density $n_s$ is defined as
\begin{equation}
    n_s=\frac{1}{\pi^2}\int_0^{k_F} dk k^2 \frac{m^*}{E^*(k)}=\frac{1}{m^*}(\varepsilon_{\rm FG}-3P_{\rm FG})~.
\end{equation}
With these above expressions (\ref{eq:PFG}) and (\ref{eq:eFG}), the scalar 
density is
\begin{equation}
    n_s=\frac{m^*}{2\pi^2}
\left[{E_F^*} k_F - {m^*}^2 \ln\left(\frac{E_F^* + k_F}{m^*} \right) \right].
\end{equation}
This analytic expression can be inserted in Eq. (\ref{eq:sigma}) which can be solved as a transcendental equation for the scalar mean field in dependence on the baryon density $n$. 
With this solution, the EoS (\ref{eq:Ph1}) and (\ref{eq:eps}) are determined and represent the EoS $P(\varepsilon)$ for the NLW model of neutron matter in parametric form.

In order to construct the transition from neutron matter to quark matter, the pressure is required as a function of the chemical potential. 
This is obtained from (\ref{eq:Ph1}) and (\ref{eq:eps}) by using the thermodynamic relation
\begin{equation}
    \mu = (P + \varepsilon) / n~.
\end{equation}

In Fig.~\ref{fig:P_mu_plots} the influence of different parameters in the quark model on the EoS is displayed and discussed.
Throughout the paper we use for characterizing the quark matter models a shorthand notation with three parameters in round brackets ($a_4$,$\Delta$[MeV],$B_{\rm eff}^{1/4}$[MeV]).
For a comparison, the pressure of pure neutron matter as a model for the hadronic phase in the core of a neutron star is described by the nonlinear Walecka (NLW) model, shown by the black dashed line.
From the relative position of the quark and neutron  matter curves, one can conclude for the possibility of a first-order phase transition by a Maxwell construction and deduce its location from the possible crossing of the curves.
As one can see in the left panel of Fig.~\ref{fig:P_mu_plots}, lowering the $a_4$ parameter stiffens the quark matter and makes a Maxwell construction impossible when the slope $dP/d\mu$ (the density) of the quark matter curve is lower than that for neutron matter. In such a case, a crossing of both curves would be unphysical, because it would describe a transition from quark matter at low chemical potentials to neutron matter at higher ones since the system has to follow the curve with the larger pressure.
A first-order phase transition described by a Maxwell construction always leads to a softening of the EoS.

The crossover construction discussed below will enforce that the system switches for increasing chemical potentials from the hadronic to the quark matter phase.
In this case it is possible to model a transition from a soft to a stiffer EoS.
In the middle panel of Fig.~\ref{fig:P_mu_plots}, we examine the variation of the diquark pairing gap $\Delta$.
For a larger gap the transition is shifted to higher densities and it is accompanied with a stiffening of the matter.
An increase in the bag constant, as shown in the rightmost panel, shifts the critical pressure of the Maxwell construction (if it is possible like in this example) to higher densities. A stiffening crossover transition, however, is moved to lower densities by the increase of the bag constant.

\begin{figure*}[!ht]
    \centering
    \subfloat{\hspace{1em}
    \includegraphics[width=.55\textwidth,trim=5em 3em 12em 6em]{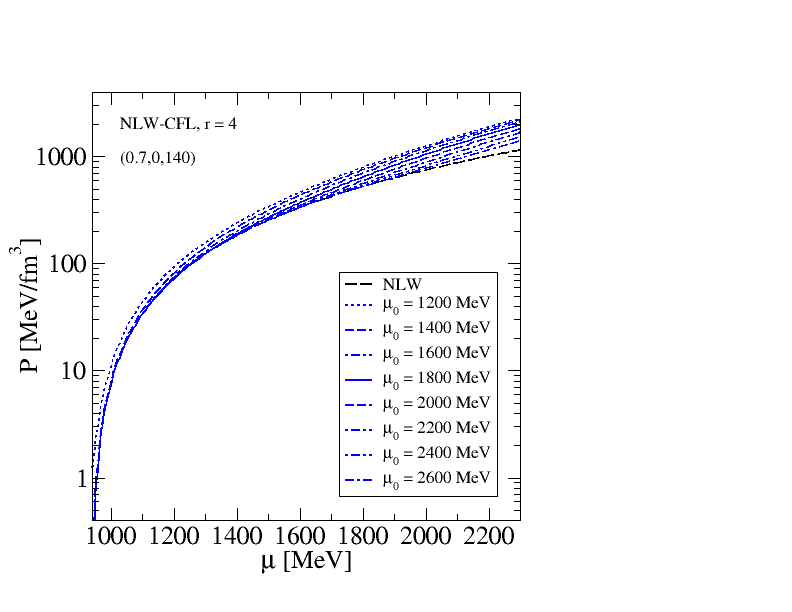}\hspace{-1.5em}
    \includegraphics[width=.55\textwidth,trim=5em 3em 12em 6em]{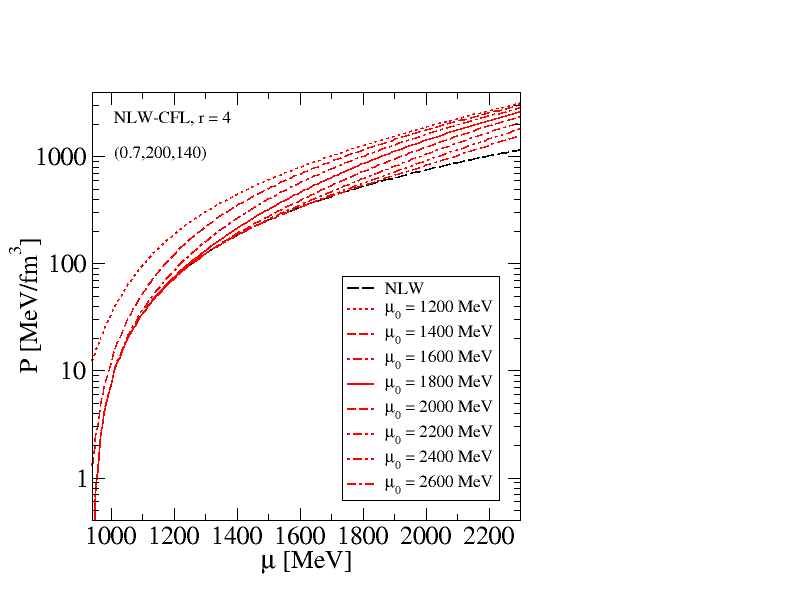}}\vspace{-0.5em}
\caption{The effect of varying the switch parameter $\mu_0$ on the pressure as a function of the baryochemical potential for $\Delta=$\,\num{0} (left panel) and $\Delta=$\,\SI{200}{MeV} (right panel). Increasing $\mu_0$ shifts the crossover transition to higher chemical potentials while a large diquark pairing gap increases the quark pressure in the hybrid EoS.}
    \label{fig:P_mu_delta200}
\end{figure*}

\subsection{Crossover EoS}

In order to construct the crossover transition from neutron matter to color superconducting quark matter, we apply the interpolation method that was introduced in 
\cite{Albright:2014gva} and applied in \cite{Kapusta:2021ney} for describing this crossover in neutron stars
\begin{equation}
    P(\mu) = S(\mu) P_q(\mu) + [1-S(\mu)]P_h(\mu)\,,
\end{equation}
whereby for the switch function we adopt the generalized form 
\begin{equation}
    S(\mu) = \exp{[-(\mu_0/\mu)^r]}~,
\end{equation}
where r=4, which was also used in \cite{Kapusta:2021ney}.
Since $P(\mu)$ is a thermodynamical potential, we can derive the other thermodynamic relations from it as
\begin{eqnarray}
    n(\mu) &=& dP(\mu)/d\mu \nonumber\\
    &=& S(\mu) n_q(\mu) + [1-S(\mu)]n_h(\mu) \nonumber\\
    &&+ 
    S^\prime (\mu) [P_q(\mu)-P_h(\mu)] 
\end{eqnarray}
and
\begin{eqnarray}
    \varepsilon(\mu) &=& -P(\mu) + \mu n(\mu) \nonumber\\
    &=& S(\mu) \varepsilon_q(\mu) + [1-S(\mu)]\varepsilon_h(\mu) \nonumber\\
    &&+ r \left( \frac{\mu_0}{\mu}\right)^{r+1} S(\mu) [P_q(\mu)-P_h(\mu)]\,, 
\end{eqnarray}
therein the relation $S^\prime (\mu) = {dS(\mu)}/{d\mu} = {r \mu_0^r}/{\mu^{r+1}\, S(\mu)} $ has been used. %

Higher order exponents ($r>$\,\num{4}) in the switch function  make the transition narrower and allow to suppress the quark matter below and the hadronic matter above the transition. 

In Fig.\,\ref{fig:P_mu_delta200} we demonstrate how  increasing the $\mu_0$ parameter in the switching function shifts the position of the crossover between hadronic and quark matter to higher chemical potentials. While in the left panel for vanishing diquark pairing gap the quark pressure is very similar to that of the NLW model for hadronic matter and the set of crossover EoS covers a rather narrow band, the increase of the quark pressure due to the large diquark gap $\Delta=$\,\SI{200}{MeV} in the right panel of Fig.\,\ref{fig:P_mu_delta200} leads to a wider band of crossover EoS.

\begin{figure*}[!ht]
 \centering  
 \subfloat{
    \hspace{-1.5em}\includegraphics[width=0.4\textwidth]{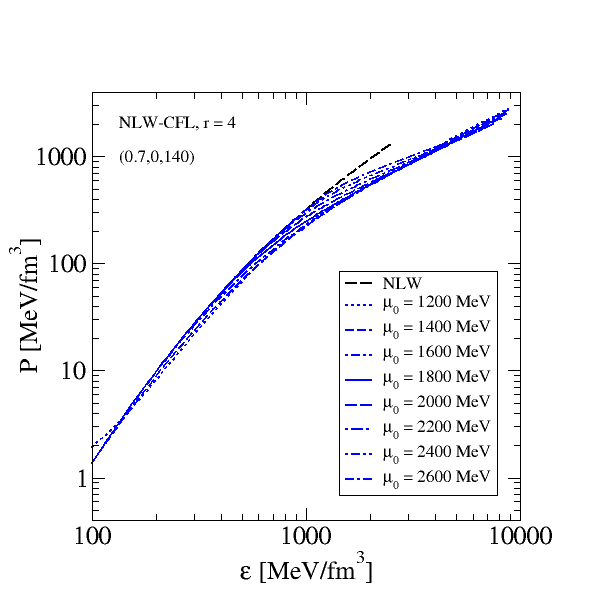}\hspace{-3.5em}
     \includegraphics[width=0.4\textwidth]{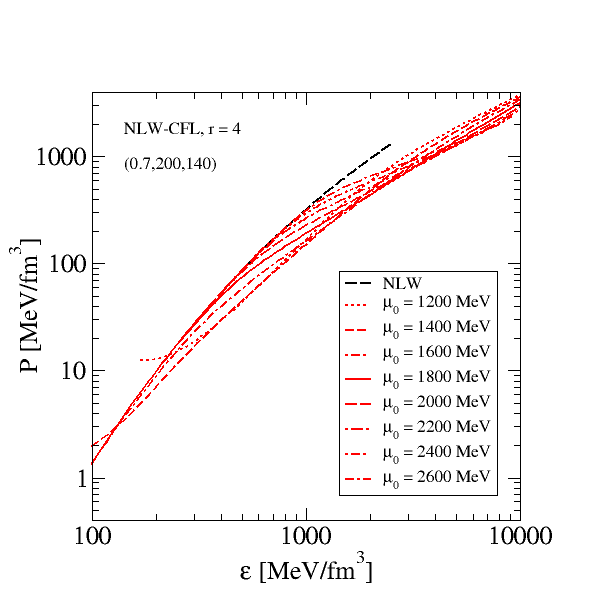}\hspace{-3.5em}
     \includegraphics[width=0.4\textwidth]{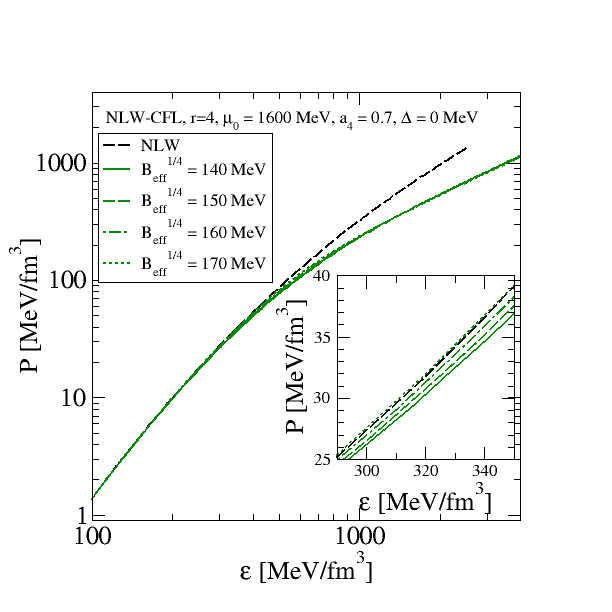}} \vspace{-4em}
     \subfloat{\hspace{-1.5em}
    \includegraphics[width=0.4\textwidth]{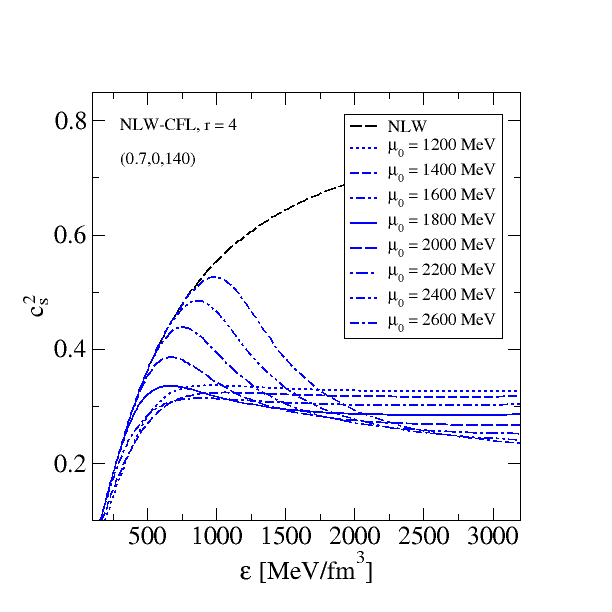}\hspace{-3.5em}
    \includegraphics[width=0.4\textwidth]{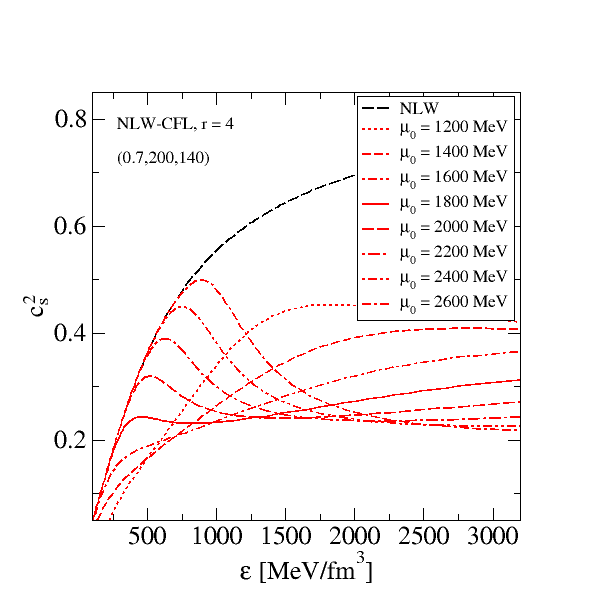}\hspace{-3.5em}
     \includegraphics[width=0.4\textwidth]{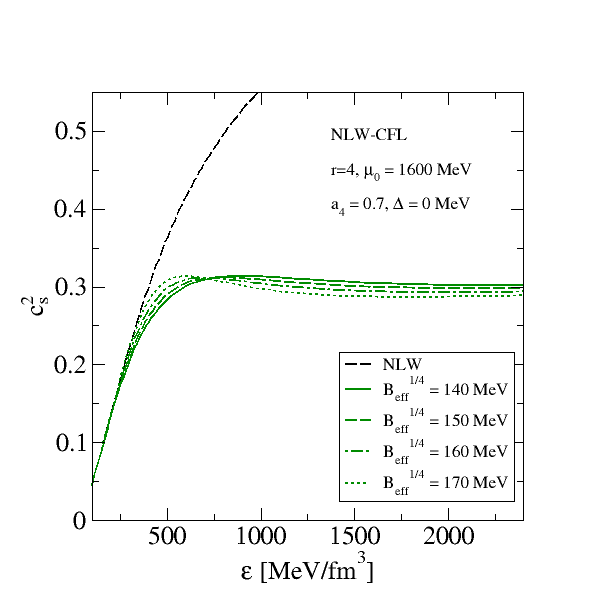}}
     \vspace{-2em}
    \caption{Upper row: Pressure as a function of the energy density for the two cases of Fig.~\ref{fig:P_mu_delta200} with vanishing diquark gap $\Delta=0$ (left panel), $\Delta=$\,\SI{200}{MeV} (middle panel) for varying switch parameter and the case $\Delta=0$ for varying bag constant $B_{\rm eff}$ at fixed $\mu_0=$\,\SI{1600}{MeV} (right panel). Lower row: Squared sound speed $c_s^2$ as a function of the energy density for the same cases as in the upper panels.}
    \label{fig:P_eps_cs2_delta0}
\end{figure*}

In Fig.~\ref{fig:P_eps_cs2_delta0} we show in the upper row three cases of parametric dependences of the pressure as a function of the energy density for which in the lower row of panels the corresponding squared sound speed is shown versus energy density.
The two leftmost cases correspond to the two panels of Fig.~\ref{fig:P_mu_delta200} with vanishing diquark gap $\Delta=0$ (left panel), $\Delta=$\,\SI{200}{MeV} (middle panel) for varying switch parameter. The case $\Delta=0$ for varying bag constant $B_{\rm eff}$ at fixed $\mu_0=$\,\SI{1600}{MeV} is shown in the rightmost panels.  
Comparing the two leftmost columns one observes that diquark pairing (color superconductivity) stiffens the EoS when the crossover transition occurs at not too high (energy) densities. For switch parameters at or above $\mu_0=$\SI{2000}{MeV}, the EoS with and without color superconductivity become indistinguishable. From the rightmost panels one observes that in the presence of the switch parameter, the variation of the bag constant has a minor effect on the crossover EoS. In a narrow domain of energy densities, the EoS switches from neutron matter to CSS quark matter behaviour.

\begin{figure*}[!ht]
 \centering 
 \subfloat{
    \hspace{-1.5em}\includegraphics[width=0.4\textwidth]{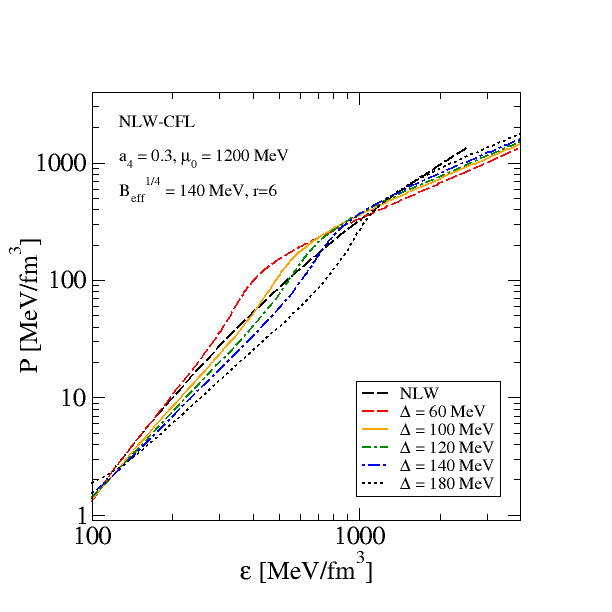}\hspace{-3.5em}
    \includegraphics[width=0.4\textwidth]{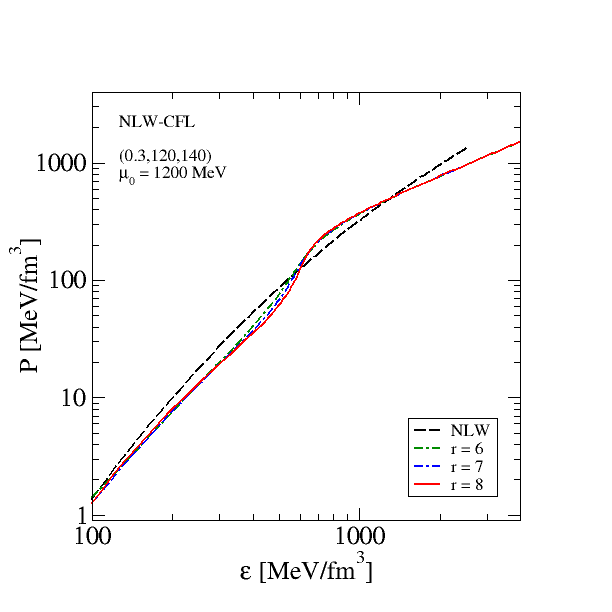}\hspace{-3.5em}
     \includegraphics[width=0.4\textwidth]{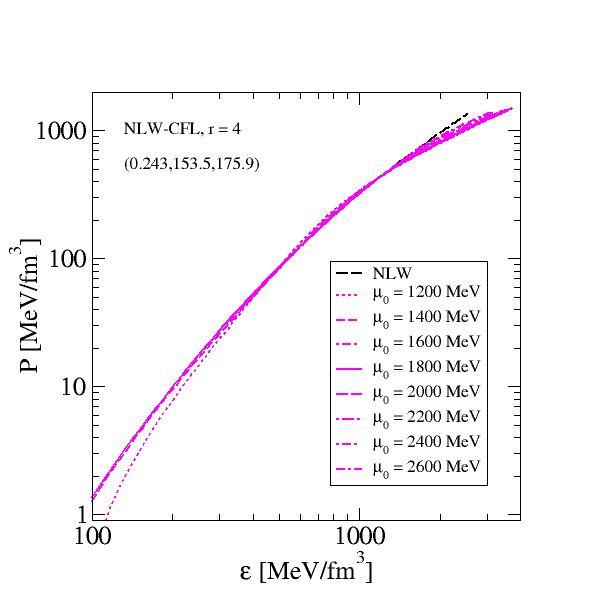}
    }\vspace{-4em}
     \subfloat{\hspace{-1.5em}
    \includegraphics[width=0.4\textwidth]{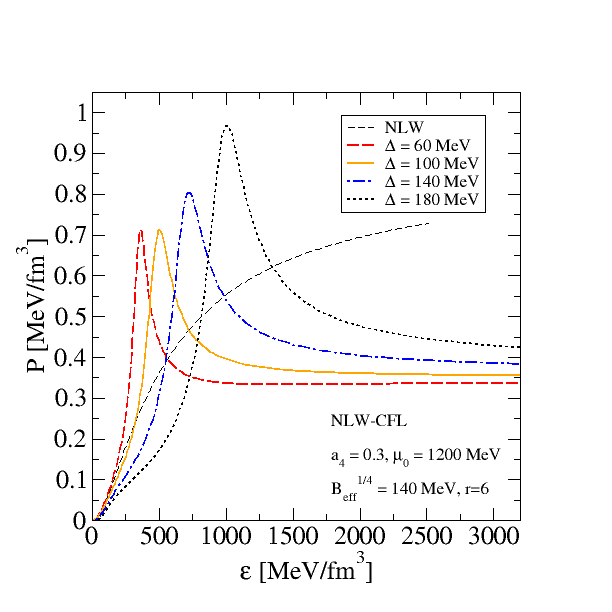}\hspace{-3.5em}
    \includegraphics[width=0.4\textwidth]{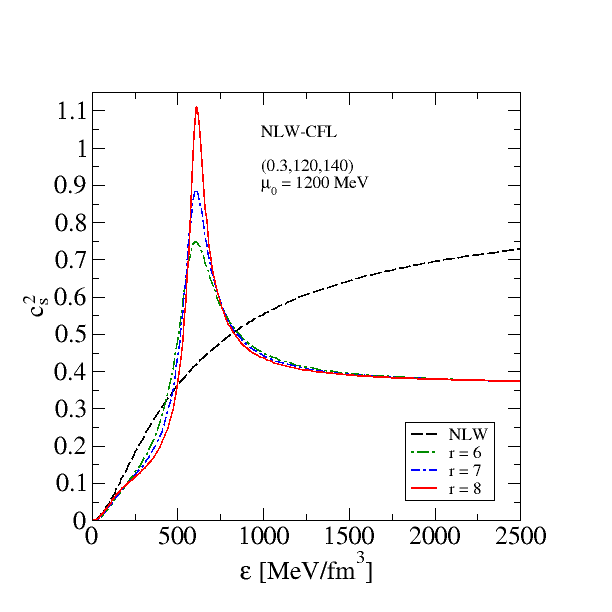}\hspace{-3.5em}
     \includegraphics[width=0.4\textwidth]{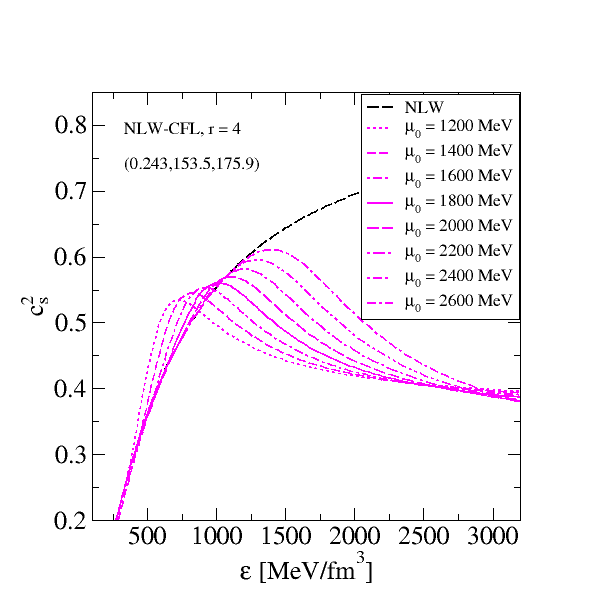}
    }\vspace{-2em}
\caption{Pressure (upper row) and $c_s^2$ (lower row) as functions of the energy density for varying diquark gap (left columns), varying width parameter $r$ of the switch function (middle column) and varying switch position parameter $\mu_0$ (right column) for strong $\alpha_s$ correction $a_4=$\,\num{0.3} (left and middle panels) and $a_4=$\,\num{0.243} (rightmost panels). For switch positions at not too high chemical potential, $\mu_0=$\,\SI{1200}{MeV} (left and middle panels) and $\mu_0<$\,\SI{2000}{MeV} (right panels), one observes a stiffening of the EoS relative to the NLW neutron matter case.}
    \label{fig:P_eps_cs2_cfl-03-delta140-r6}
\end{figure*}

In Fig.~\ref{fig:P_eps_cs2_cfl-03-delta140-r6}, we investigate the effect of varying the diquark gap (left columns), varying the width parameter $r$ of the switch function (middle column) and varying the switch position parameter $\mu_0$ (right column) on the pressure (upper row) and the squared sound speed (lower row) as functions of the energy density for strong $\alpha_s$ correction $a_4=$\,\num{0.3} (left and middle panels) and $a_4=$\,\num{0.243} (rightmost panels). For switch positions at not too high chemical potential, $\mu_0=$\,\SI{1200}{MeV} (left and middle panels) and $\mu_0<$\,\SI{2000}{MeV} (right panels), one observes a stiffening of the EoS relative to the NLW neutron matter case.

Generally, a switching function can always be applied. 
However, there is the question if such a crossover transition describes a physical or an unphysical crossing.
An unphysical crossing is the case when at low densities (before the crossing) the quark matter pressure is above the hadronic one and vice-versa at high densities (after the crossing).
If such an unphysical crossing appears, it is better to use a replacement interpolation method than a mixing (see\,\cite{Abgaryan:2018gqp} for a comparison), as it has been discussed in\,\cite{Ayriyan:2021prr} and\,\cite{Baym:2017whm}. A characteristic feature of such a crossover construction is that it results in a stiffening of the EoS, which is a feature very welcome for modern neutron star phenomenology\,\cite{Masuda:2012ed}. 

There are several approaches to that problem.
In the easiest one defines upper and lower boundaries of the transition 
region outside of which the respective EOS can be trusted and inserts an interpolating function\,\cite{Ayriyan:2021prr} in between\,\footnote{In 
the generalization to the QCD phase diagram at finite temperatures, the transition region of the EoS can either be replaced (for example by an Ising model) or interpolated in a corridor between the bordering hadronic and quark matter phases.
Such a transition corridor allows also to insert a CEP with the characteristic critical exponents in its vicinity
\cite{Nonaka:2004pg,Parotto:2018pwx,Kapusta:2021oco}. 
On this basis the effects of a CEP on the phenomenology (e.g., the hydrodynamical evolution of a heavy-ion collision and related observables) can be studied.}.

\subsection{Calculation of astrophysical observables}

From the equations of state, we derive possible neutron star radii and masses. 
These can directly be compared to observations from the combined observations by NICER and XMM Newton of the millisecond pulsar J0740+6620 according to the analysis of Miller \emph{et al.}\,\cite{Miller:2021qha}
Additionally, the tidal deformability $\Lambda$ can be calculated for the considered sequence of neutron star masses and be compared 
to the constraint obtained from the gravitational wave signal that was observed for the binary neutron star merger GW170818\,\cite{LIGOScientific:2018cki} in the mass range $M\approx 1.4~M_\odot$. 
To evaluate the neutron star properties one has to solve the Tolman-Oppenheimer-Volkoff (TOV) equations for a static non-rotating, spherical-symmetric star\,\cite{PhysRev.55.364,PhysRev.55.374}
\begin{align}
    \frac{dP(r)}{dr} &= \frac{G(\epsilon(r))+P(r))(M(r)+4\pi r^3 P(r))}{r(r-2GM(r))}\\
    \frac{dM(r)}{dr} &= 4\pi r^2 \epsilon(r)\,,
    \label{eq:TOV}
\end{align}
with $P(r=R)=0$ and $P(r=0)=P_c$ as boundary conditions for a star with mass $M$ and radius $R$.
The astrophysical observables were calculated using the code by Anton Motornenko~\cite{amotornenko2018}.
For all solutions of the TOV equations with the hybrid neutron star EoS the crust EoS by Baym, Pethick and Sutherland (BPS)\,\cite{baym1971} has been added.

\begin{figure*}[!ht]
 \centering 
 \subfloat{\hspace{-1.5em}
    \includegraphics[width=0.4\textwidth]{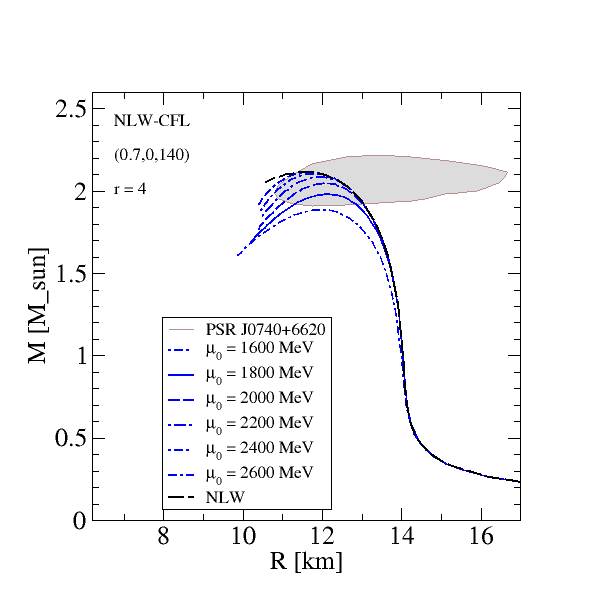}\hspace{-4em}
     \includegraphics[width=0.4\textwidth]{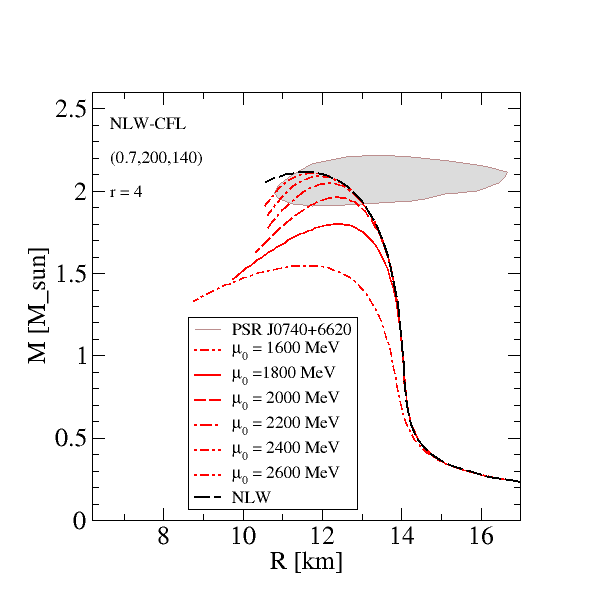}\hspace{-4em}
      \includegraphics[width=0.4\textwidth]{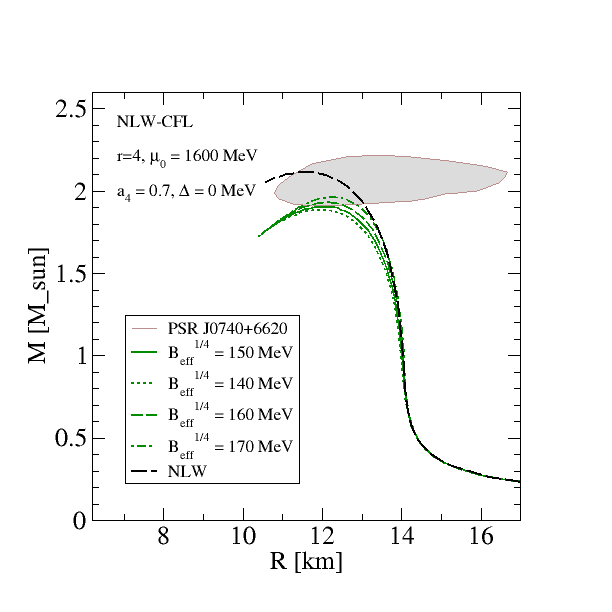}
    }\vspace{-4em}
 \subfloat{   \hspace{-1.5em}
    \includegraphics[width=0.4\textwidth]{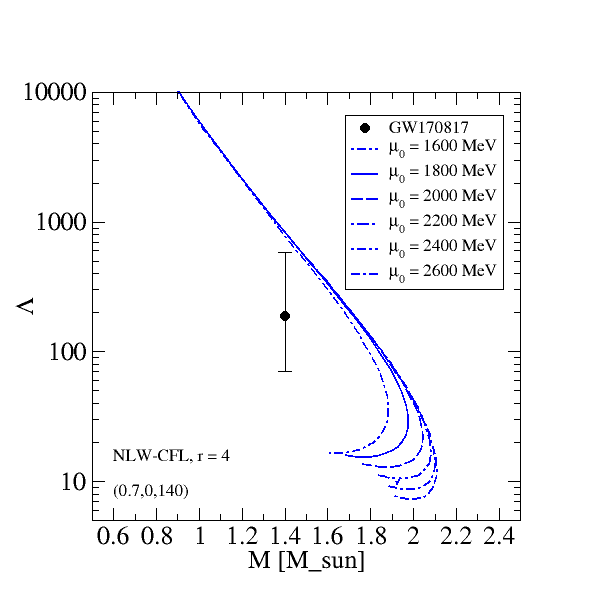}\hspace{-4em}
    \includegraphics[width=0.4\textwidth]{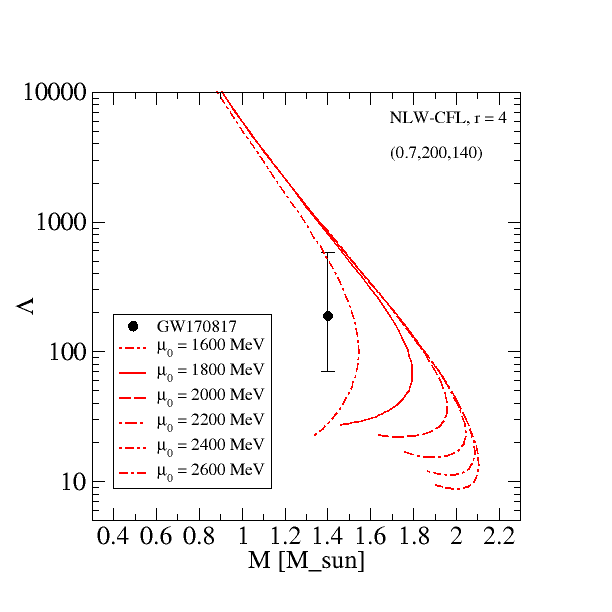}\hspace{-4em}
    \includegraphics[width=0.4\textwidth]{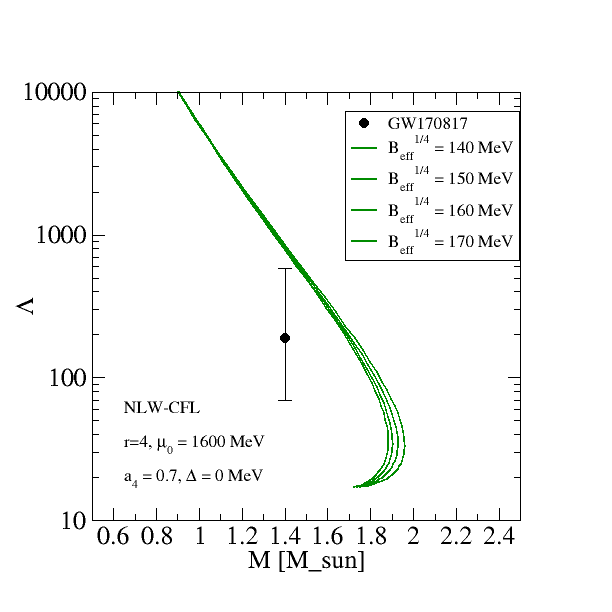}
    }\vspace{-2em}
    \caption{Mass vs. radius (upper panels) and tidal deformability vs. mass (lower panels for the EoS cases discussed in Fig.~\ref{fig:P_mu_delta200}. None fulfills simultaneously both observational constraints, from the combined observations by NICER and XMM Newton of the millisecond pulsar J0740+6620 according to the analysis of Miller\,\emph{et al.}\,\cite{Miller:2021qha} shown as the grey hatched region in the upper panels and from the tidal deformability of GW170817 as reported in\,\cite{LIGOScientific:2018cki}.}
    \label{fig:NS_delta0}
\end{figure*}

\begin{figure*}[!ht]
 \centering 
 \subfloat{\hspace{-1.5em}
    \includegraphics[width=0.4\textwidth]{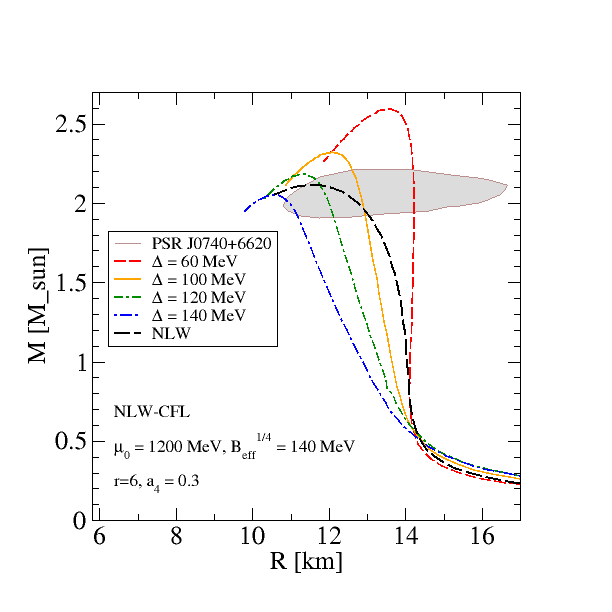}\hspace{-4em}
    \includegraphics[width=0.4\textwidth]{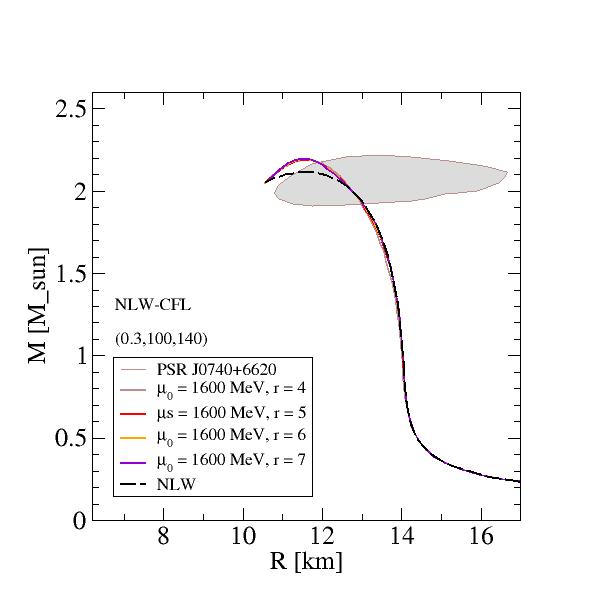}\hspace{-4em}
    \includegraphics[width=0.4\textwidth]{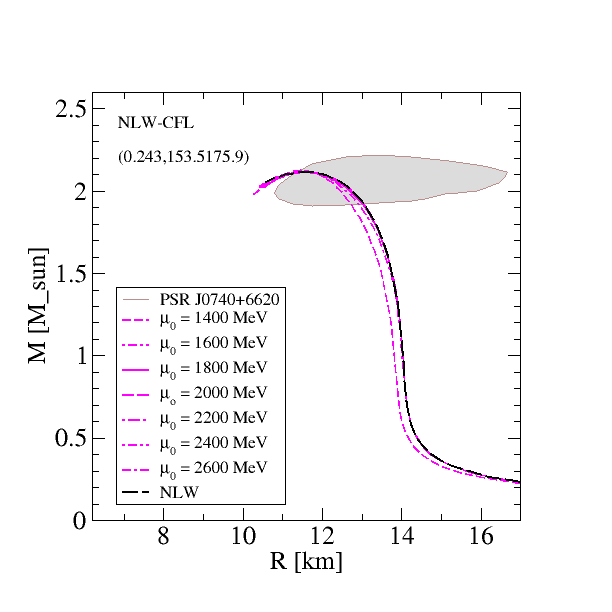}
    }\vspace{-4em}
    \subfloat{\hspace{-1.5em}
    \includegraphics[width=0.4\textwidth]{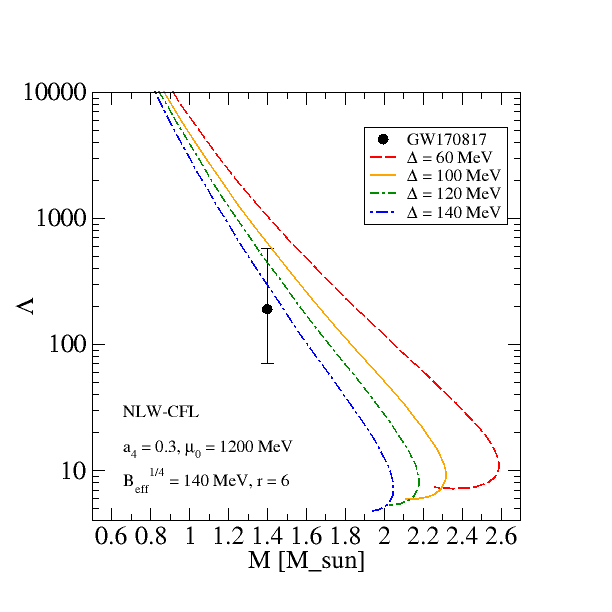}\hspace{-4em}
    \includegraphics[width=0.4\textwidth,]{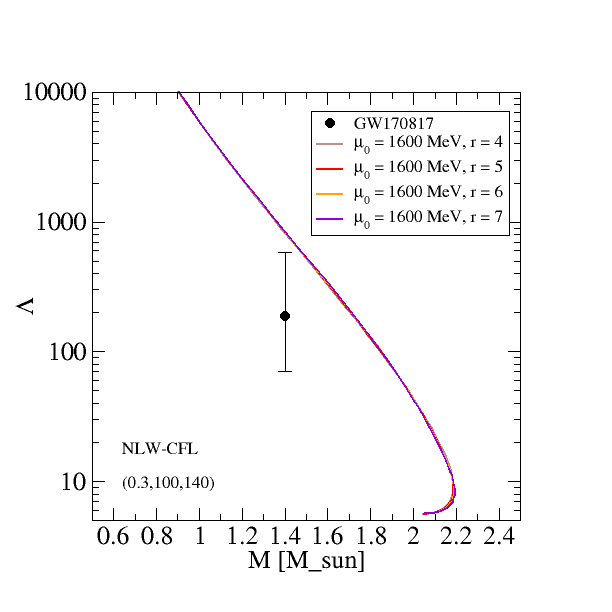}\hspace{-4em}
    \includegraphics[width=0.4\textwidth]{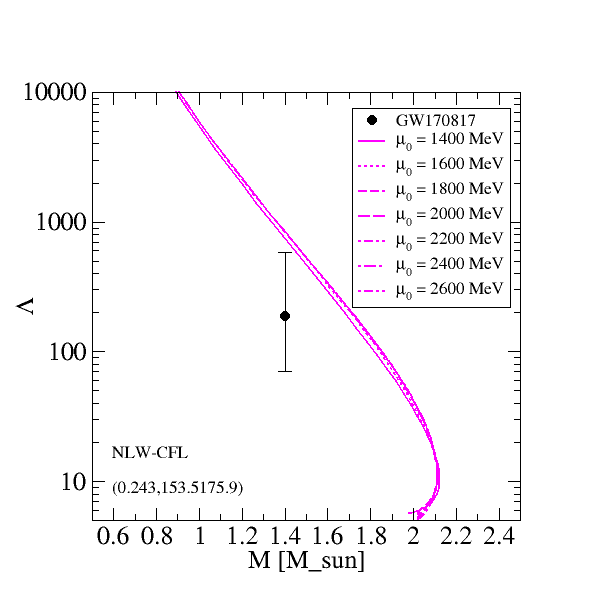}
    }\vspace{-2em}
    \caption{Like Fig.~\ref{fig:NS_delta0}, but for the EoS cases discussed in Fig.~\ref{fig:P_eps_cs2_cfl-03-delta140-r6}. Only the color superconducting model with sharp crossover parameter $r=$\num{6} and sufficiently large gap ($\Delta=$\SIlist{100,120}{MeV}) at low crossover position $\mu_0=$\SI{1200}{MeV} can fulfill both constraints from observation simultaneously.}
    \label{fig:NS_cfl03-100-140-r47}
\end{figure*}

\section{Results}

For the parametrization of the NLW model we use the values given in the textbook by Glendenning \cite{Glendenning:1997wn}, the effective mass $m^*/m=0.8$, the incompressibility of symmetric nuclear matter at saturation $K=$\,\SI{250}{MeV} and the asymmetry energy at saturation $E_s=$\,\SI{32}{MeV}. With these phenomenological data the parameters $m_\sigma=$\,\SI{550}{MeV}, ${g_\sigma}/{m_\sigma}=$\,\num{8.692}, ${g_\omega}/{m_\omega}=$\,\num{4.0243}, ${g_\rho}/{m_\rho}=$\,\num{4.4369}, $b=$\,\num{8.898E-3} and $c=$\,\num{7.708E-3}.
The parameter of the switch function is chosen as $\mu_0=$\,\num{1400},\,\num{1500},\,\num{1600},\,\SI{1800}{MeV}.

It is interesting to consider the squared sound speed as a quantity related to the stiffness of the EoS, applying Eq.~(\ref{eq:cs2}). 
Finally, we use the crossover EoS to solve the TOV equations for stellar structure and obtain the corresponding mass-radius and tidal deformability-mass sequences that can be compared with observational data, see figures \ref{fig:NS_delta0} and \ref{fig:NS_cfl03-100-140-r47}. 

We distinguish two classes of parametrizations. 
The first class has an $\mathcal{O}(\alpha_s)$ correction to the quark matter EoS based on the standard one-loop running of the QCD $\beta$-function that results in a moderate stiffening of the EoS with a typical value of $a_4=$\,\num{0.7}.
This class was considered in \cite{Kapusta:2021ney} and it would allow a traditional Gibbs construction of the phase transition, with values for $\Delta$ and $B_{\rm eff}$ that would balance each other. 
The second class uses much lower values for the $a_4$ parameter that could be motivated by a nonperturbative enhancement of the strong coupling and a corresponding stiffening of the quark matter EoS to a degree that exceeds the stiffness of the nonlinear Walecka model EoS and would entail an unphysical crossing in the $P-\mu$ plane if one were to attempt a Gibbs construction.
Typical values are $a_4=$\,\num{0.3} or lower.
One recognizes this class of crossover EoS by the effect that their sound speed exceeds that of the NLW model at energy densities that are typical for compact star interiors, see lower panels of Fig.\,\ref{fig:P_eps_cs2_cfl-03-delta140-r6}.
Such a behaviour of the squared sound speed is typical for the microscopic realization \cite{McLerran:2018hbz}
of the quarkyonic matter hypothesis \cite{McLerran:2007qj}
in which the unifying concept of quark and hadron matter is realized according to which in high-density matter baryons populate the surface of a Fermi sea of quarks.

From the comparison with the recent mass-radius data of the massive pulsar PSR J0740+6620\,\cite{Miller:2021qha} that were obtained by the Maryland-Illinois team of the NICER collaboration, we may conclude that for the first class of models a too small value for the switch function parameter $\mu_0<$\,\SI{1400}{MeV} could be excluded, because it would 
lead to a too low maximum mass of pulsars which is excluded by observation, see also\,\cite{Fonseca:2021wxt}.
A similar conclusion has been drawn by Kapusta and Welle\,\cite{Kapusta:2021ney}, but with a different value for the limiting $\mu_0$ parameter.
This may be attributed to the fact that in our work we allow for color superconductivity in the quark matter phase which has an influence on the stiffness of the neutron star matter in the relevant region that determines the maximum mass of pulsars. 
However, when considering in addition the tidal deformability constraint from GW170817 \cite{LIGOScientific:2018cki}, we have found no parametrization of the quark matter model and the switching function that would simultaneously fulfill both constraints from neutron star phenomenology.

For the second class of models, however, we have found a parametrization that would fulfill both, tidal deformability 
and mass-radius constraints. It corresponds to $a_4=$\,\num{0.3}, a diquark pairing gap of $\Delta=$\,\SI{120}{MeV} and $B_{\rm eff}^{1/4}=$\,\SI{140}{MeV} with a low switch function parameter 
$\mu_0=$\,\SI{1200}{MeV}, see Fig.\,\ref{fig:P_eps_cs2_cfl-03-delta140-r6}. A narrowing of the transition region by an increased switch function exponent $r=$\,\num{6} helps to avoid a modification of the resulting crossover EoS at low densities, where it should remain in accordance with the known  properties at nuclear saturation as parametrized in the NLW model.

\section{Conclusions}
We have performed a reanalysis of the switch function parameters for a unified description of quark-hadron matter with a crossover transition by employing modern mass-radius and tidal deformability constraints from multi-messenger astronomy. 
We find that for a simultaneous description of these constraints it is essential to enrich the pQCD ansatz for the quark matter EoS with nonperturbative aspects. 
These are a low $a_4$ parameter pointing to a nonperturbative enhancement of the strong coupling at low energies, a nonvanishing diquark pairing gap indicating the color superconducting state of quark matter and an effective bag constant for confining effects. 
Moreover, we find it reasonable to narrow the transition by using a larger exponent $r=$\num{6} than in \cite{Kapusta:2021ney}
and favor a lower value of the switch function parameter $\mu_0=$\SI{1200}{MeV}.
The influence of such a low crossover on the stiffness of the EoS is limited. 
In general, one should be very careful when using such switching functions. 

The present work uses the simplifying assumption of a pure neutron matter EoS in the hadronic phase and massless quarks in the quark matter phase so that no leptons in beta-equilibrium needed to be considered. In a more realistic study, these assumptions should be relaxed.
In order to obtain the favorable crossover EoS in the multi-parameter model presented here, one should invoke a Bayesian analysis \cite{Ayriyan:2018blj,Blaschke:2020qqj,Ayriyan:2021prr}.
The approach using the switching function should be contrasted to the methods that employ an interpolating (polynomial) function within fixed limits for the density \cite{Abgaryan:2018gqp,Blaschke:2018pva,Ayriyan:2021prr}, 
to obtain the crossover transition. 

\begin{acknowledgments}
The authors thank Oleksii Ivanytskyi for help with the nonlinear Walecka model EoS and Michal Marczenko for his assistance in solving the TOV equations.
The work of D.B. has been supported in part by the Polish National Science Centre (NCN) under grant No. 2019/33/B/ST9/03059 and by the Russian Foundation for Basic Research (RFBR) under grant No. 18-02-40137.
He was supported by the Russian Federal Program "Priority-2030".
The authors acknowledge the COST Action CA16214 "PHAROS" for supporting their networking activities, 
in particular their particcipation in the PHAROS Training School on "Equation of State of Dense Matter and Multi-Messenger Astronomy" in Karpacz, June 13-19, 2021.
This work is part of a project that has received funding from the European Union's Horizon 2020 research and innovation program under the grant agreement STRONG - 2020 - No 824093.
\end{acknowledgments}

\bibliography{qcp}

\end{document}